# MASTER ROBOTIC NET

**Vladimir Lipunov[1], Victor Kornilov[1], Evgeny Gorbovskoy[1], Nikolaj Shatskij[1], Dmitry Kuvshinov[1], Nataly Tyurina[1], Alexander Belinski[1], Alexander Krylov[1], Pavel Balanutsa[1], Vadim Chazov[1], Artem Kuznetsov[1], Petr Kortunov[1], Anatoly Sankovich[1], Andrey Tlatov[2], A.Parkhomenko[2], Vadim Krushinsky[3], Ivan Zalozhnyh[3], A.Popov[3], Taisia Kopytova[3], Kirill Ivanov[4], Sergey Yazev[4], Vladimir Yurkov[5]**

[1] *Moscow State University, Sternberg Astronomical Institute, 119991, 13, Univeristetskij pr-t, Moscow, Russia*
[2] *Kislovodsk Solar Station, 357700 p.o. Box 145, 100, Gagarina st., Russia*
[3] *Ural State University, 620083, 51, Lenina pr-t, Ekaterinburg, Russia*
[4] *Irkutsk State University, 664003, 1, Karl Marks ul., Irkutsk, Russia*
[5] *Blagoveschensk State Pedagogical University, 675000, 104, Lenina st., Blagoveschensk, Amur reg., Russia*

The main goal of the MASTER-Net project is to produce a unique fast sky survey with all sky observed over a single night down to a limiting magnitude of 19—20mag. Such a survey will make it possible to address a number of fundamental problems: search for dark energy via the discovery and photometry of supernovas (including SNIa), search for exoplanets, microlensing effects, discovery of minor bodies in the Solar System and space-junk monitoring. All MASTER telescopes can be guided by alerts, and we plan to observe prompt optical emission from gamma-ray bursts synchronously in several filters and in several polarization planes.

## 1. Introduction

The MASTER project (Mobile Astronomical System of TElescope Robots) was launched in 2002, when the first Russian robotic telescope was installed near Moscow on an automated mounting with automatic roof, weather station, and alert system (Lipunov et al., [1,2,3]). By early 2008 we posted about 100 GCN circulars, recorded optical emission from three gamma-ray bursts, and discovered four supernovas (Lipunov et al., [4], Tyurina et al.[5]). Unfortunately, very poor astroclimatic conditions near Moscow made it impossible to demonstrate the potential of our telescope to its full extent. However, we obtained about 80 000 images, each with a size of six square degrees, and used them to successfully develop a software pipeline, which allows us not only to perform in real time the extraction of images and their astrometric and photometric reduction, but also to automatically classify the sources identified --- i.e., determine which type of astronomical objects --- supernovas, minor planets, comets, man-made satellites, meteors, or optical transients they belong to (Fig.1, 2). Moreover, our facility can also be used for «alert-driven» observations of optical emission of gamma-ray bursts (Tyurina et al.[5]).

Our experience with the operation of robotic telescopes allowed us to develop the wide-field MASTER optical facility with optimum parameters, which can be used to address a wide range of astrophysical tasks. We actually developed a multipurpose optical facility, which is similar in composition to the one independently proposed by Bogdan Paczynski (Paczynski, [6]). The first MASTER telescopes and their individual components are already installed at three sites in Russia – near Kislovodsk (Caucasian Mountain Observatory of the Sternberg Astronomical Institute of the Moscow State University), in Urals (Kourovka Observatory of Ural State University), and near Irkutsk (Irkutsk State University Observatory). In summer 2009 we also plan to deploy similar facilities near Irkutsk and Blagoveshchensk.

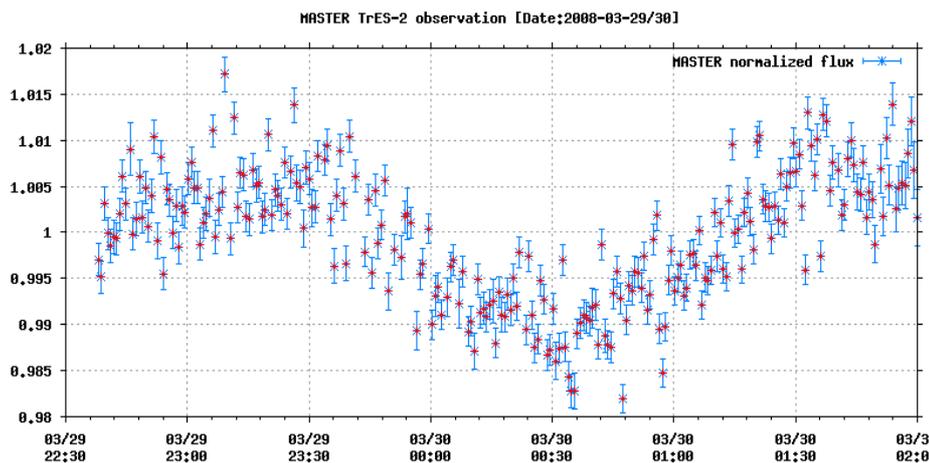

**Figure 1:** Folow-up exoplanet transit observations by MASTER (Moscow)

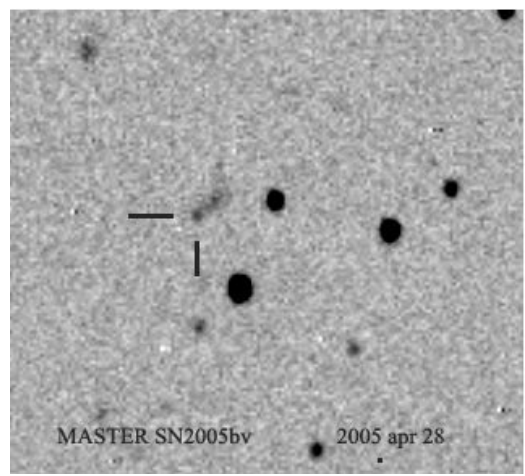

**Figure 2:** The first supernova discovered in Russia (SN2005bv)

## 2. MASTER Wide-Field Robotic Facility.

The MASTER wide-field robotic telescope that we developed is manufactured by OAO Optika Moscow Association. The facility consists of two instruments: MASTER II wide-field optical telescope ("Santel 400") and MASTER very wide-field (VWF) cameras.

## 2.1. MASTER II Wide-Field Robotic Telescope.

A single fast German equatorial mounting carries two 1:2.5 focal-ratio symmetrical high-aperture 400-mm catadioptric telescopes (Fig. 3). The telescope has extra degree of freedom, which allows its tubes to be aligned or misaligned, making it possible to double the size of the field of view and, when operating in the alert mode (with collinear tubes), to perform synchronous photometry of rapidly-varying objects in standard broad-band filters (B,V,R,I) and their polarimetry at different polarizations (Fig.4).

Table 1 lists the technical parameters of MASTER II telescope. The telescope is equipped with a shell-type automatic shelter, which, when open, allows the entire sky to be observed. This facility is also equipped with a time synchronization system, a weather station, and a cloud cover sensor.

**Table 1.** Composition and specification of MASTER II telescope..

| Item | Parameter | Number | Remark | Special name |
|---|---|---|---|---|
| Telescope (2 pieces) | Diameter | 400 mm | Light-beam diameter | Hamilton system (Santel 400) |
| | Focus | 1 000 mm | | |
| | Weight | 50 kg | On each side | |
| | Field of view | 2 x 4 square degrees | For an Apogee AltaU16 type CCD camera | |
| Mounting | Weight capacity | Up to 100 kg | German | NTM |
| | Maximum positioning speed | 30 deg/s | | |
| Dome | Diameter | 3.6 m | | |
| | Weight | 500 kg | | |
| 2 CCD cameras | Number of pixels | 16 Mpix | | |
| | Pixel size | 9 microns | | |
| | Image scale | 1.84 " per pix | | |
| Cloud Sensor | | | The Boltwood Cloud Sensor II | |
| GPS | Accuracy of | 100 | | |

|  | synchronization | microseconds |  |  |
| --- | --- | --- | --- | --- |
| Server | RAM | 8Gb |  |  |
|  | Processor | Xion 54XX |  |  |
|  | HDDs | Rate 10Tb |  |  |
| Survey rate |  | 480 sq. deg/h |  |  |
| Limiting magnitude for 1-min exposure |  | 19m | under optimum astroclimatic conditions |  |

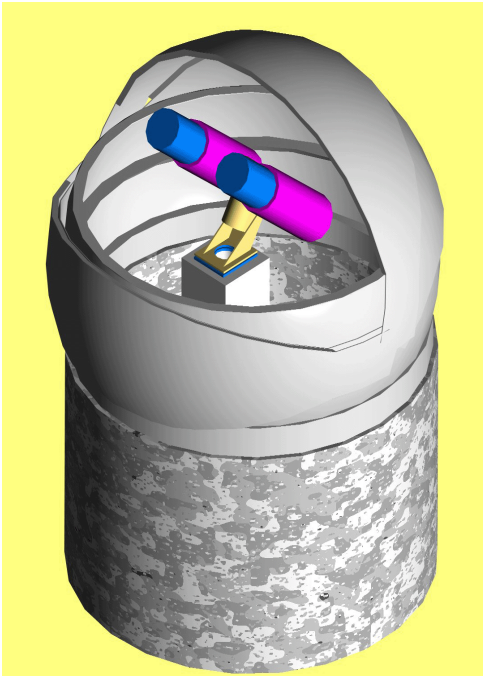

**Figure 3:** MASTER II wide-field robotic telescope (manufactured by OAO Optika Moscow Association). Each tube is equipped by a photometer with a set of B,V,R,I filters and polarizers.

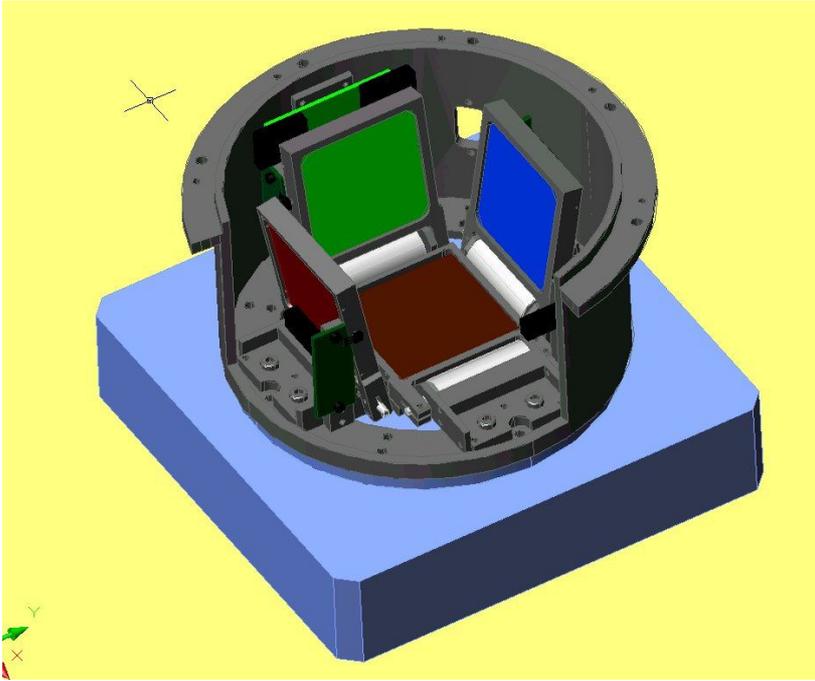

**Figure 4:** Photometer with filters (3) and polarimeter (1) based on an Apogee AltaU16M CCD camera.

**2.2 MASTER VWF very wide-field camera.**

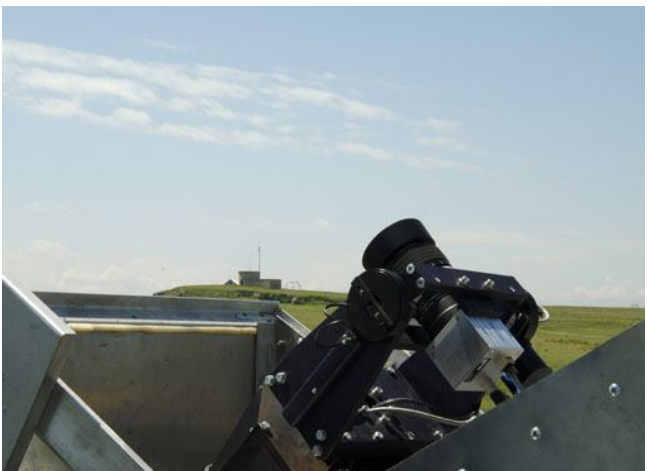

**Figure 5:** MASTER Very Wide Field camera

V.G. Kornilov developed a robotic very wide-field MASTER VWF camera with a maximum field of view of 1000 square degrees (Figure 5) for synchronous observations of optical emission from gamma-ray bursts. The facility uses a fork parallactic capable of carrying 2 CCD cameras with up to 10-cm aperture lenses. The facility has an autonomous high-quality roof designed by N.I. Shatskii. Our experience of its operation under extreme conditions of Russian winter (with temperatures as low as -40 degree Celsius) demonstrates good survival potential of all systems. Table 2 lists the specifications of the very wide field camera. The cameras are adapted to all large-aperture Nikor and Nikor-Ziess lenses. Our facility uses two lenses with the focal distances of 50 and 85 mm. The cameras allow continuous sky imaging with a minimum exposure of 150 milliseconds.

**Table 1.** Composition and specification of MASTER VWF camera.

| Item | Parameter | Value | Remark | Special name |
|---|---|---|---|---|
| Lens (2 pieces) | Diameter | 35.7mm | Light-beam diameter | Nikor 50 f/1.4 |
| | Focus | 50 mm | | |
| | Field of view | 1040 square degrees | For a CCD camera with a chip size of 24 x 36 mm | |
| Lens (2 pieces) | Diameter | 60.7mm | Light-beam diameter | Zeiss-Nikor 85/1.4 |
| | Focus | 85 mm | | |
| | Field of view | 432 square degrees | | |
| Mounting | Weight capacity | Up to 10 kg | Fork | Frictional |
| | Maximum positioning speed | 6 deg/s | | |
| | Positioning accuracy | 1 arcmin | | |
| Dome | Diameter | 1x1.5 m | | |
| | Weight | 30 kg | | |
| CCD cameras (2 pieces) | Number of pixels | 12 Mpix | | |
| | Pixel size | 9 microns | | |
| | Image scale | 36" per pix<br>20" per pix | for f50<br>for f85 | |
| Limiting magnitude for 1-min exposure | | $11^m$<br>$13^m$ | for f50<br>for f85 | |

**3. Geographic Location of the Stations of the MASTER Network as of May 2009.**

The locations of station sites were determined by two factors: (1) sites must be about ~2 hours apart in longitude and (2) sites must provide minimum required infrastructure (power supply, Internet, warm, warm room for servers). Such sites can usually be found at university observatories (Fig.6).

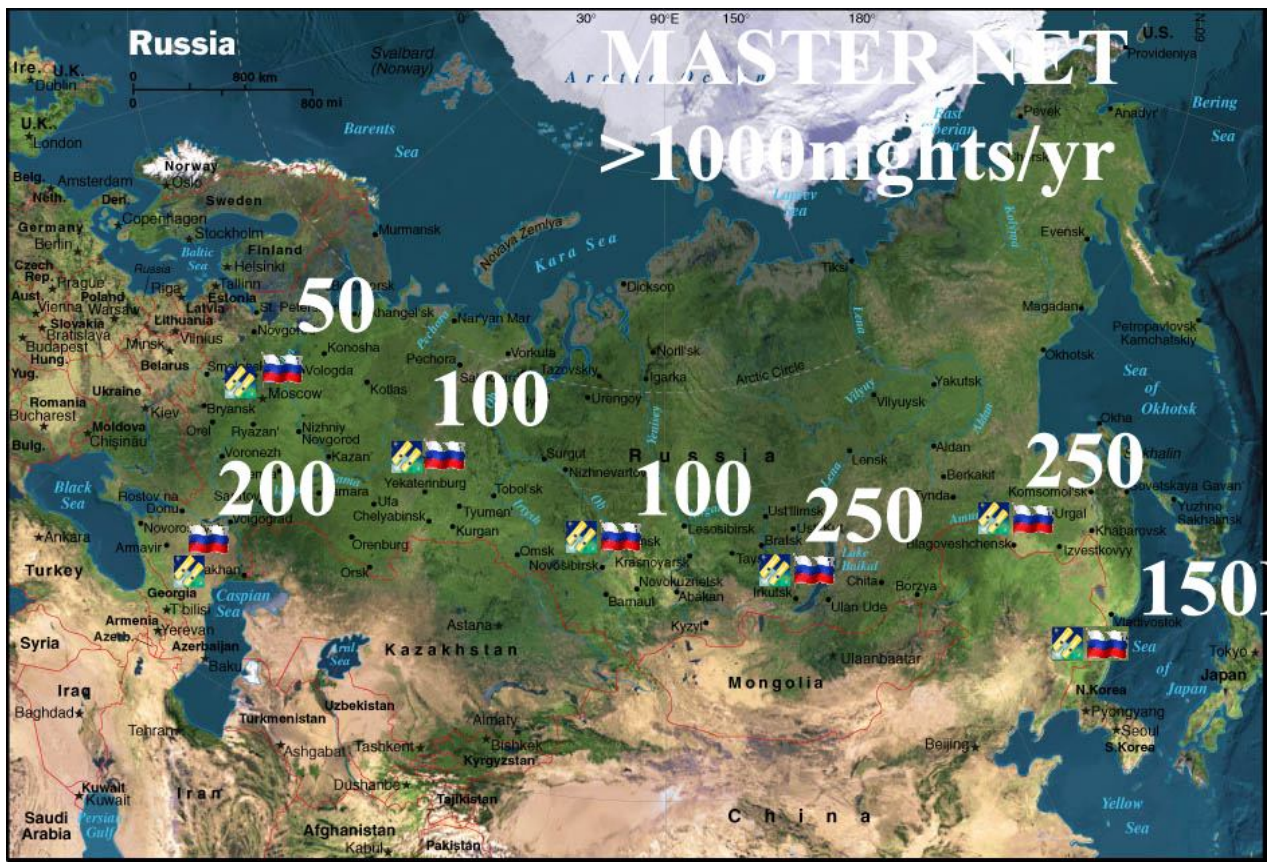

**Figure 6:** Russian robotic network MASTER. The flags indicate the planned and already equipped sites of network stations. From West to East: Moscow (MASTER I), Kislovodsk (MASTER II, MASTER VWF-4), Ekaterinburg (MASTER II), Novosibirsk (planed), Irkutsk (MASTER VWF-2), Blagoveshchensk (planned), and Ussuriisk (planned). The average annual number of clear nights is indicated near each site.

The participants to the project already include Moscow State University (Fig.7), Ural State University (Fig.8), Irkutsk State University, and Blagoveshchensk Pedagogical University. The network extends over seven hours in longitude, thereby ensuring virtually continuous 24-hour observations during winter seasons.

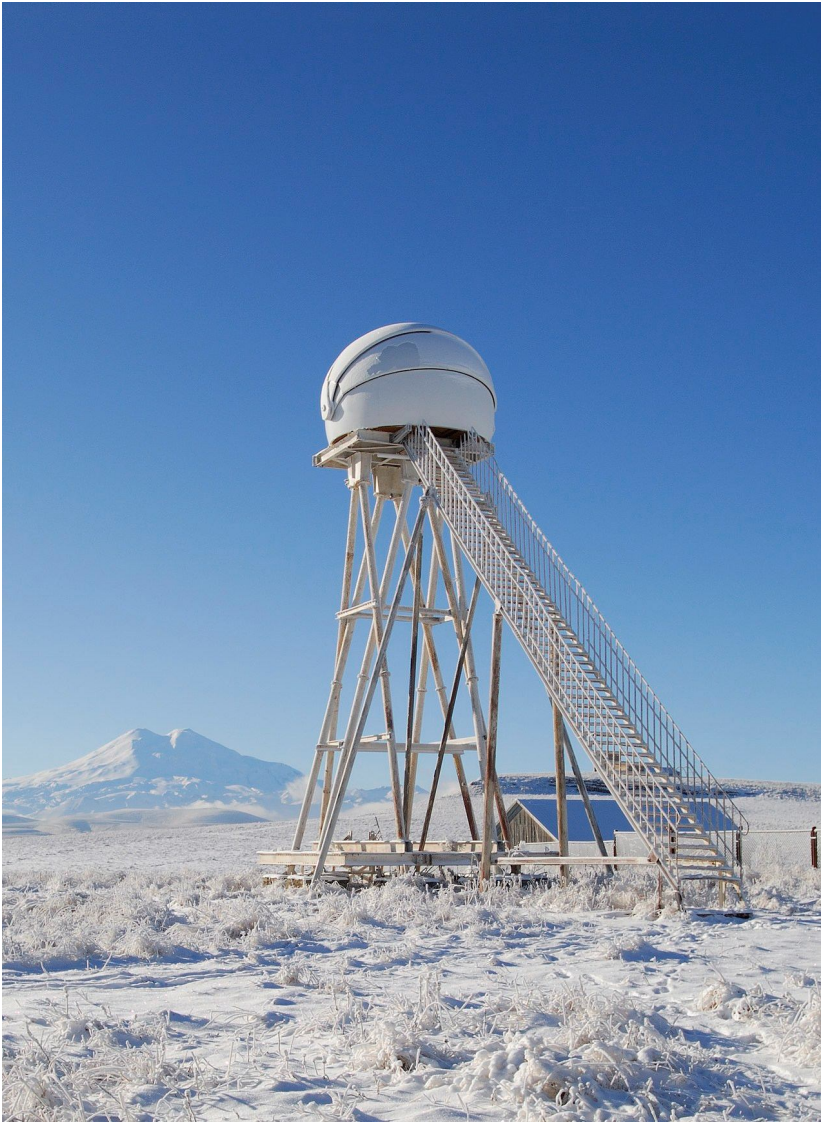

**Figure 7**. MASTER II (Kislovodsk, Sternberg Astronomical Institute, Moscow State University). Photo by A.Belinskiy.

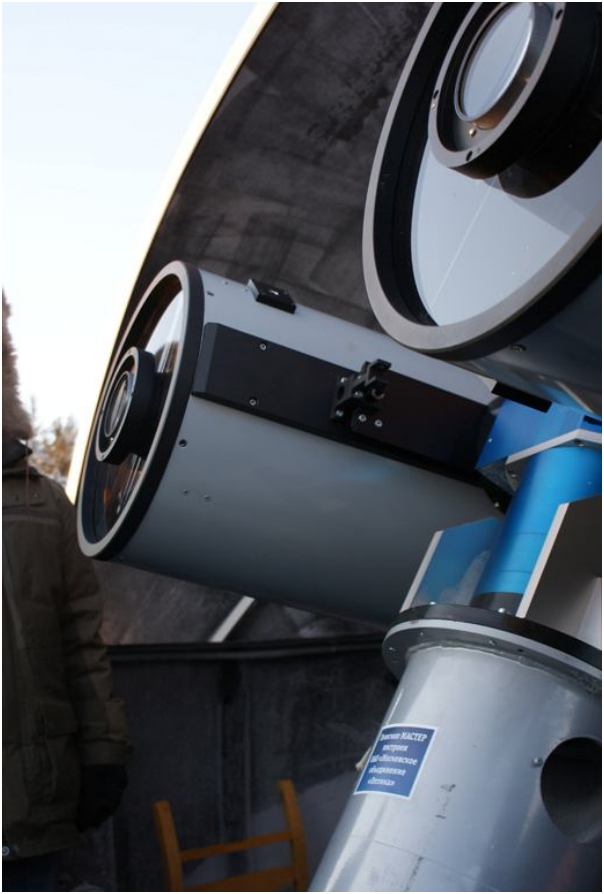

**Figure 8.** MASTER II (Kourovka, near Ekaterinburg, Ural State University).

When deployed, the first step MASTER II telescope network will have the following parameters:
- Total field of view   - 38 square degrees
- Survey rate  - 2000 square degrees per hour down to a limiting magnitude of 19-20.
- Alert positioning speed – up to 30 degrees/s

## 3. Expected scientific results.

### 3.1 GRB mystery.

Observations of the prompt emission of gamma-ray bursts, which is determined by the physical processes in the central engine, are of fundamental importance for understanding the nature of these bursts (Lipunov and Gorbovskoy, [7,8]).

First and foremost, we expect that MASTER II telescopes will allow us to obtain multicolor photometric and polarimetric follow up observations for alerts from gamma-ray space observatories. We also plan to continue synchronous observations with very wide-field cameras. Recall that have performed five synchronous observations of gamma-ray error boxes since November 2008 (Gorbovskoy et al., [9]).

### 3.2 Dark energy.
The first mass observations of type Ia supernovas showed (Reiss A.G. et al., [10]; Perlimutter S. Et al., [11]) that the Universe expands with acceleration, which may be due to the so-called dark energy. This result follows from observations of several dozen supernovas. On the other hand, present-day search surveys miss most of the supernovas. The first computations of the velocities of explosions of cosmological supernovas performed by Jorgensen et al., [12] showed that the number of supernovas exploding in our Universe with magnitudes brighter than **m** can be described by the following formula:

$$N(<m) \sim 10^5 \, 10^{3/5(m-20)} \, yrs^{-1}, \, (\, 15 < m < 22)$$

Note that without the correction for host-galaxy extinction $20^m$ corresponds to a redshift of $z \sim 0.2$-$0.3$, where the effect of cosmic energy of vacuum becomes quite appreciable.
This actually means that MASTER can discover several thousand supernovas annually and perform multicolor photometry of these objects.

### 3.3. Other tasks.
A complete survey of the entire visible sky during a single night will allow to perform a number of tasks with no special observing plan (provided, naturally, proper coordination between the stations of the network):
- **Search for orphan flares**
- **Search for exoplanets**
- **Search for dangerous asteroids**
- **Search for microlensing effects**
- **Search for and monitoring of space waste.**

ACKNOWLEDGMENTS

The authors thank the General Director of the "OPTIKA" Association S.M. Bodrov for providing the MASTER project with necessary expensive equipment.

## 4. References.